\documentclass{aa}
\usepackage{graphicx}
\begin{document}
   \title{Using color photometry to separate transiting exoplanets from false positives}

   \author{B. Tingley \inst{1}}

   \offprints{B. Tingley}

   \institute{$^{1}$Research School of Astronomy and Astrophysics, ANU
                    Cotter Road, Weston, Canberra ACT 2611, Australia
              \email{tingley@mso.anu.edu.au}
             }

   \date{Received X, 2003; accepted Y, 200Z}

   \abstract{The radial velocity technique is currently used to classify
transiting objects. While capable of identifying grazing binary eclipses,
this technique cannot reliably identify blends, a chance overlap of a
faint background eclipsing binary with an ordinary foreground star.
Blends generally have no observable radial velocity shifts, as the
foreground star is brighter by several magnitudes and therefore
dominates the spectrum, but their combined light can produce events
that closely resemble those produced by transiting exoplanets.

The radial velocity technique takes advantage of the mass difference
between planets and stars to classify exoplanet candidates. However,
the existence of blends renders this difference an unreliable
discriminator. Another difference must therefore be utilized for
this classification -- the physical size of the transiting body. Due
to the dependence of limb darkening on color, planets and stars produce
subtly different transit shapes. These differences can be relatively
weak, little more than 1/10th the transit depth. However, the presence
of even small color differences between the individual components
of the blend increases this difference. This paper will show that
this color difference is capable of discriminating between exoplanets
and blends reliably, theoretically capable of classifying even
terrestrial-class transits, unlike the radial velocity technique.

   }

   \maketitle
%

\section{Introduction}

Of all of the difficulties facing the search for transiting exoplanets,
the issue of certainty is hardest to resolve. There are many phenomena
capable of producing events that appear very similar to planetary
transits. It is an observational challenge to separate the exoplanets
from "false positives" -- the non-planetary sources that manifest
low-amplitude transit-like events. Currently, radial velocities
are measured for exoplanetary candidates to make this discrimination.
These measurements are quite capable of identifying grazing eclipsing
binaries, which are the primary source of false positives. Beyond
this, however, they are of limited use. They cannot, for example, classify
exoplanets significantly smaller than Jupiter -- currently, the least
massive exoplanet discovered by the radial velocity technique is 0.2
M$_J$. This limitation is caused by the natural asteroseismic radial
velocity oscillations present in main sequence stars, which mask the
radial velocity shifts. Moreover, the need for ultra high-resolution
spectra to detect the radial velocity shifts limits the depth to which
exoplanets can be verified. Even the best facilities in the world can
only go so deep before it becomes impossible to attain the necessary
spectral resolution to identify even short-period Jovian planets.
These effects have already impacted efforts to discover exoplanets
and will be a major obstacle for ambitious projects
such as Kepler and COROT, which aim to discover terrestrial exoplanets.
Without any means to classify the vast numbers of very shallow
transiting systems consistent with exoplanets that are significantly
less than Jupiter mass, it will be difficult to achieve the science
goals of the missions, to identify planets and estimate the fraction
of stars with planets.

However, the most serious failing of radial velocities is their inability
(in most cases) to identify "blends", a chance superposition
where the light from a star is contaminated by that of a fainter
eclipsing binary along the same line of sight. In this circumstance,
the depth of the eclipse
is reduced by the presence of the bright ordinary star, possibly to
the point that it is consistent with an exoplanetary transit. To
compound this problem, the combined light of a blend is often dominated
by the brighter star, hiding the radial velocity shift of the
faint binary. In some circumstances, high-resolution spectra can reveal
blends through the presence of strong line asymmetries or extra spectral
lines rather than radial velocity variations (Konacki et al. \cite{konacki}).
Another recently published technique proposes using observed transit
parameters to identify blends, effectively taking advantage of the
change in shape of the transit between blends and exoplanets (due to
limb darkening and size differences) to make the identification
(Seager \& Mall\'e n-Ornelas \cite{seager}). It can be difficult
to measure the shape of the transit accurately enough for classification
purposes, however.

The OGLE-III campaign (Udalski et al. \cite{udalski1}, \cite{udalski2}),
the most successful planet hunt to date, demonstrated
the various limitations of the radial velocity technique. Of the 59
candidates that were identified during the 2001 campaign of OGLE-III,
7 were considered too faint for
follow-up observation and only 6 of the remaining candidates had radial
velocity variations less than a few $km s^{-1}$ and no sign of secondary
transits or out-of-transit variations. Of these 6, one was clearly
a blend, one clearly a planet, and the other 4 remain unclassified.
This means that out of the 46 that were truly interesting candidates, 11
could not be classified using radial velocities, despite the use
use of world-class facilities Keck and VLT to make the observations
(Konacki et al. \cite{konacki}, Dreizler et al. \cite{dreizler}).

Clearly, the radial velocity technique will be insufficient for the needs of
satellites such as Kepler and COROT. These projects will not go as deep
as OGLE-III, but can be expected to observe a large number of
transit-like events of all depths, both Jovian-class (corresponding to a giant planet)
and terrestrial-class (corresponding to a terrestrial planet).
Clearly, the success of these projects is crucially dependent on the
ability to classify exoplanet candidates reliably.

Another method for classifying transits has appeared many times
in the literature over the past 35 years but has never received a full
development. It was first mentioned by Rosenblatt (\cite{rosenblatt}),
then again some years later by Borucki et al. (\cite{borucki1}), and
recently by Drake (\cite{drake}) and Charbonneau (\cite{charbonneau}),
who is involved in a group assembling a telescope dedicated to the
use of this technique for classifying planet candidates (Kotredes et
al. \cite{kotredes}).
It involves observing the color change of the system during transit --
most exoplanet transits produce a very distinctive color index "signature"
that is clearly distinguishable from a grazing binary or a blend.
The strength of this signature in $B-I$ is about 15\% of the depth of
the transit in the individual colors and will usually be even
stronger for blends. Only in the unlikely occurrence that all three
stars in the blend have essentially the same $B-I$ will a blend have
a signature weaker than this.

These differences between exoplanet transits and binary or blended
transit are potentially of great import in the search for exoplanets,
allowing for a better classification of candidates. The sensitivity of
blended transits to color difference only increases its utility. This
article will explore these differences and demonstrate their
practicality. First, the method for modeling the transits and
constructing the blends will be discussed. The $\delta(B-I)$ and
$\delta(V-I)$ of exoplanets and blends will then be compared,
exploring differences in color between the components. The results will
demonstrate that this method is a viable technique for discriminating
between blends and exoplanets.

\section{Modeling the transits and constructing blends}

Modeling transits is a basic application of generic limb darkening laws.
Complex limb darkening laws exist, but the degree to which they match
reality is poorly known. It is an observational challenge to measure
limb-darkening, especially for solar-type (and later) stars. The limb
darkening for only one such star (besides the Sun) has been observed,
from a high-magnification
microlensing event (Abe et al. \cite{abe}). In addition, metallicity has
an effect on limb darkening and is in most cases an unknown.
For these reasons, it is sufficient for the purposes of this analysis
to use a simple limb darkening law, especially as only the lower main
sequence is of concern and only first order precision can be expected
regardless. The limb darkening law used, taken from Astrophysical
Quantities (Allen \cite{allen}), is

\begin{equation}
\frac{I'_{\lambda}(\theta)}{I'_{0}(\theta)} = 1 - u_1(1 - \cos \theta)
\end{equation}

\noindent
where $u_1$ is estimated for the $B$, $V$ and $I$ filters, using Astrophysical
Quantities and weighted for the passbands from Bessell (\cite{bessell}).
The constants derived are: 0.75 for $B$, 0.55 for $V$ and 0.4 for $I$.
Of course, these values are for the Sun and there are inevitably changes
along the main sequence. However, the characteristic of this limb darkening
law that makes this technique possible is preserved -- the increased
concentration of blue light to the center of the disk compared to the
red light, evidenced by the fact that $u_1$ is almost twice as large 
for $B$ as it is for $I$. 

To construct a blend, one must take the results from a modeled binary
eclipse and its colors and combines them with a third star with the proper
separation to produce a blended transit with the desired depth in the
chosen color. A blend effectively consists of three components that
affect the color change during transit: a constant component and two
eclipsing components. The constant component can exist of any number of
stars along the line of sight, and can be physically in front of or
behind the eclipsing components -- or even both. Ultimately, it is only
the compare colors and magnitude of the constant component and the eclipsing
components, plus the depth of the unblended eclipse in the various colors
and the separation of the components that matter. This method of
constructing blends is otherwise completely general.

A relationship for determining the proper separation can be
derived from the modeled depth of the binary eclipse and the desired
depth of the blend using the basic relation between magnitudes and fluxes:

\begin{equation}
m_1 - m_2 = -2.5 \log \frac{f_1}{f_2}
\end{equation}

\noindent
where $m_1$ and $m_2$ are the magnitudes of the two components to be
compared and $f_1$ and $f_2$ are their corresponding fluxes. This
equation is then applied twice: once to the blend outside eclipse and
at eclipse maximum and again to the constant component and the
eclipsing component outside of eclipse. With this, the following
equation can be derived:

\begin{equation}
m_e = m_c - 2.5 \log \left(\frac{1-10^{-0.4D}}{10^{-0.4D}-d}\right)
\end{equation}

\begin{figure*}
   \centering
   \includegraphics{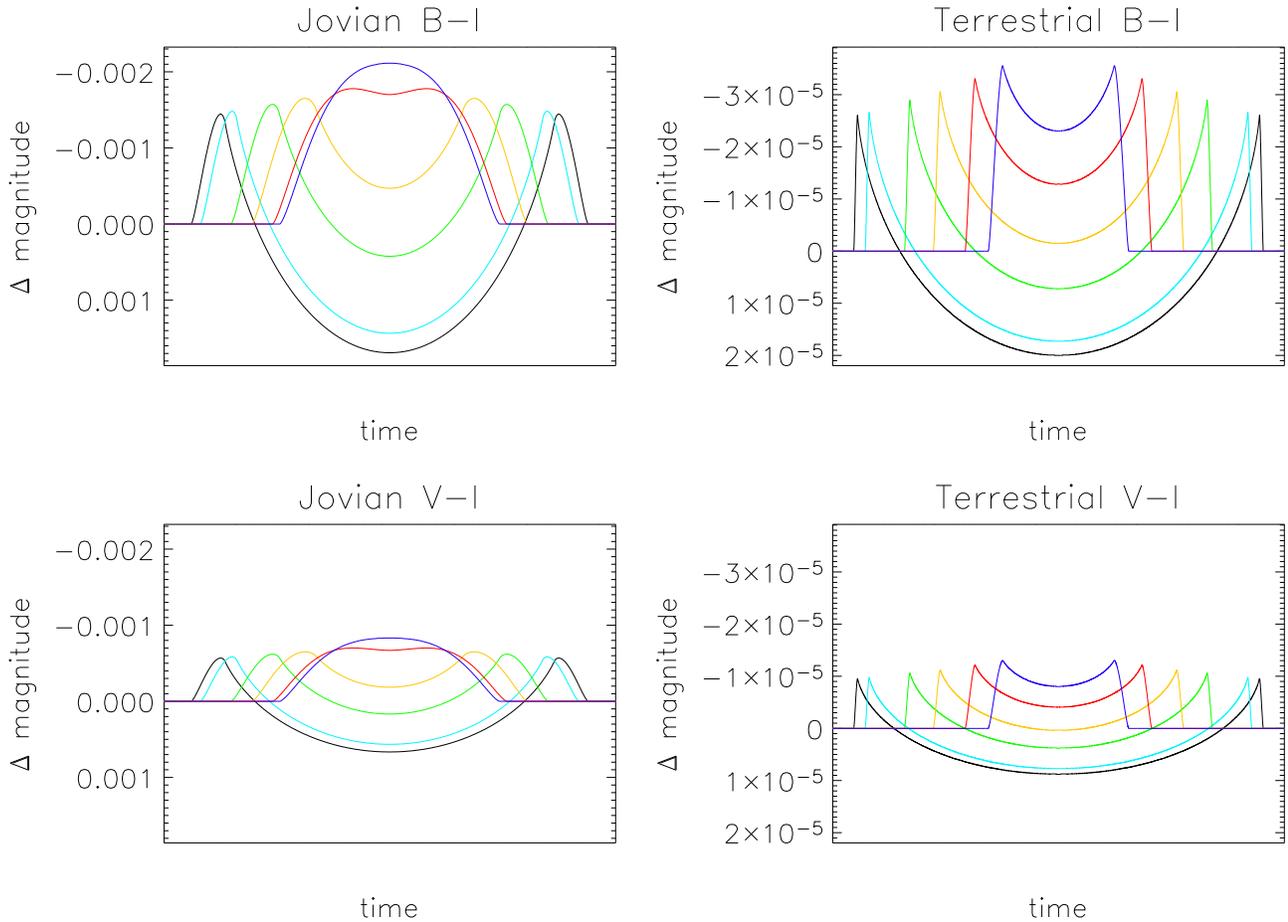}
   \caption{This figure shows modeled Jovian (depth = 0.01 magnitudes in I)
and terrestrial (depth=0.0001 magnitudes) transits for different projected
inclinations in both $B-I$ and $V-I$. The projected inclinations used were
0 (narrowest), $\frac{R}{3}$, $\frac{2R}{3}$, $\frac{4R}{5}$, $\frac{9R}{10}$
and $\frac{19R}{20}$ (narrowest).
The double-horned profile is typical of exoplanetary transits -- stellar
blends cannot produce it. However, grazing Jovian transits without the
double-horned profile can be mimicked by blends. The size of the transiting
planet must be increased for increasing projected inclinations
to maintain a transit depth in $I$.}
\end{figure*}

\noindent
where $m_e$ is the total magnitude of the eclipsing component outside
of transit, $m_c$ is the magnitude of the constant component, $D$ is the
desired depth of the blended transit in magnitudes and $d$ is the
fractional depth of the binary eclipse. This can be extended for magnitudes
in a single filter, but then the transit depths in that filter must be used.

\section{Analysis}

The analysis in this paper uses transit depths of the blends in $I$ and
considers two different classes transit depths, Jovian (depth = 0.01
magnitudes) and terrestrial (depth = 0.0001 magnitudes). It explores the
effect of color differences between the constant component and the
eclipsing components. It also explores differences in color between the
eclipsing components. The most likely case where all three components of
the blend will be different colors is a logical extension of the studied
cases. Beyond stellar blends, there is the possibility that a transiting
Jovian exoplanet might blend with other stars to create a terrestrial-class
transit. The effects of color on this hypothetical blend are also explored.

Due to the nature of the analysis, it is only the color differences
between the components that is significant. The actual absolute colors
of the components involved is irrelevant. So, while a spectral type of G2 was
used to make the figures, any spectral type could have been used and the
results would have been the same, aside from minor differences in the
limb darkening laws and radii of exoplanets used in order to model
transit depths.

\section{Results of transit analysis}

Figure 1 shows the $B-I$ and $V-I$ for Jovian and terrestrial exoplanets
at different projected inclinations, defined as the closest approach
of the center of the secondary to the center of the primary during transit.
The sizes of the modeled planets were adjusted slightly so that the
transits would have the same depth in $I$ -- 0.01 magnitudes for Jovian
planets and 0.0001 magnitudes for terrestrial.

Both terrestrial and Jovian exoplanets create transits which have double
horned profiles for most projected inclinations. These horns are
more than twice as high in $B-I$ as they are in $V-I$
and are sharper for terrestrial exoplanets than they
are for Jovian. They occur because of the small size of the exoplanets
compared to the size of the star, which accentuates the general property
of limb darkening that photospheres are more centrally concentrated
at shorter wavelengths. This means that as the exoplanet crosses the limb,
the star appears more blue, as red light is preferentially occulted. As
the exoplanet moves across the disk of the star, the star appears more red
as the exoplanet completely clears the limb of the star (allowing the 
occulted red light in again) and begins preferentially occulting blue light.

\begin{figure*}
   \centering
   \includegraphics{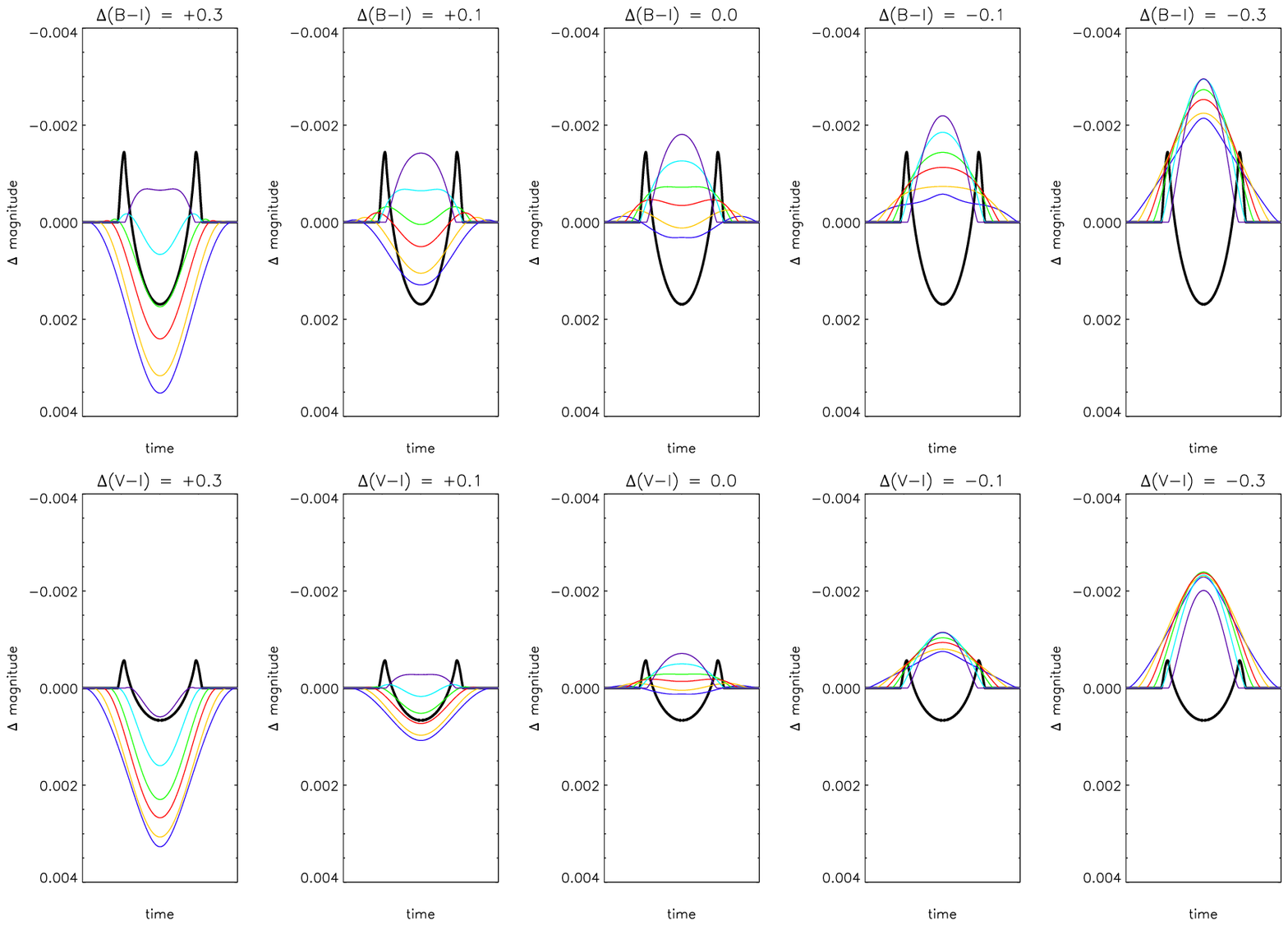}
   \caption{This figure shows blended Jovian-class transits in instances
where the constant component has different colors than the eclipsing
components, which are identical. The thin lines depict different projected
inclinations for the
given color difference. The projected inclinations used are $\frac{R}{3}$
(broadest), $R$, $\frac{4R}{3}$, $\frac{3R}{2}$, $\frac{5R}{3}$ and $1.78 R$
(narrowest). The heavy black line is a central (i = 0) Jovian transit
shown for comparison. Further diluting
the light to form a terrestrial-class blend will not change the shape of
the blended events, only their depth, so the corresponding modeled 
terrestrial-class events are not included.}
\end{figure*}

In Figure 1, the height of these horns appears to increase with projected
inclination. This
is however a result of the fact that these transits are modeled to have
a fixed depth in $I$. Maintaining this requires an increase in exoplanet
radius, as transits away from central are not as strong, which in turn
increases the height of the horns. If the exoplanet radius had been fixed,
the horns would be the same height for all projected inclinations
where the exoplanet crosses
the limb of the star entirely, as that is that point where the horn occurs.
Obviously, the horn should have the same height under these conditions,
as it occurs the same exoplanet/star configuration -- the exoplanet lying
just over the limb of the star.

Figure 2 shows a series of plots that depict the $B-I$ and $V-I$ of blends
constructed of identical eclipsing components with constant components of
slightly different colors for several different projected inclinations.
For these blends, the separation between the eclipsing and constant
components was adjusted such that there
was a fixed transit depth in the combined emission, as described in
Equation 2. It is important to realize that changes in $B-I$ and $V-I$ can be
treated independently -- the shape of the transits in $B-I$ is set by
the $\Delta(B-I)$ between the constant and eclipsing components and
independent of any differences in any other color band.

The thick line superimposed over these plots is a modeled central
Jovian transit of the same depth. By comparing figures 1 and 2, one
can see the the worst case scenarios involve grazing Jovian transits,
which lack a double-horned profile. It is possible to construct a blend
in this case that would mimic a giant planet to the point that they
would be indistinguishable. However, it would require that the
eclipsing component had the same $V-I$ as the constant component while
being slightly bluer (the $\Delta(B-I)$s shown is Figure 2 are for the
constant component) and moreover be at least moderately grazing.
This is helpful, as grazing eclipses are swallower, requiring that
that the separation between the two components must be reduced to
compensate. As the density of background stars increases rapidly
with magnitude, this decreases the probability that a configuration that
produces a Jovian transit-like event could occur while increasing the
chance that its nature might be betrayed by spectra taken for radial
velocity measurements.

\begin{figure*}
   \centering
   \includegraphics{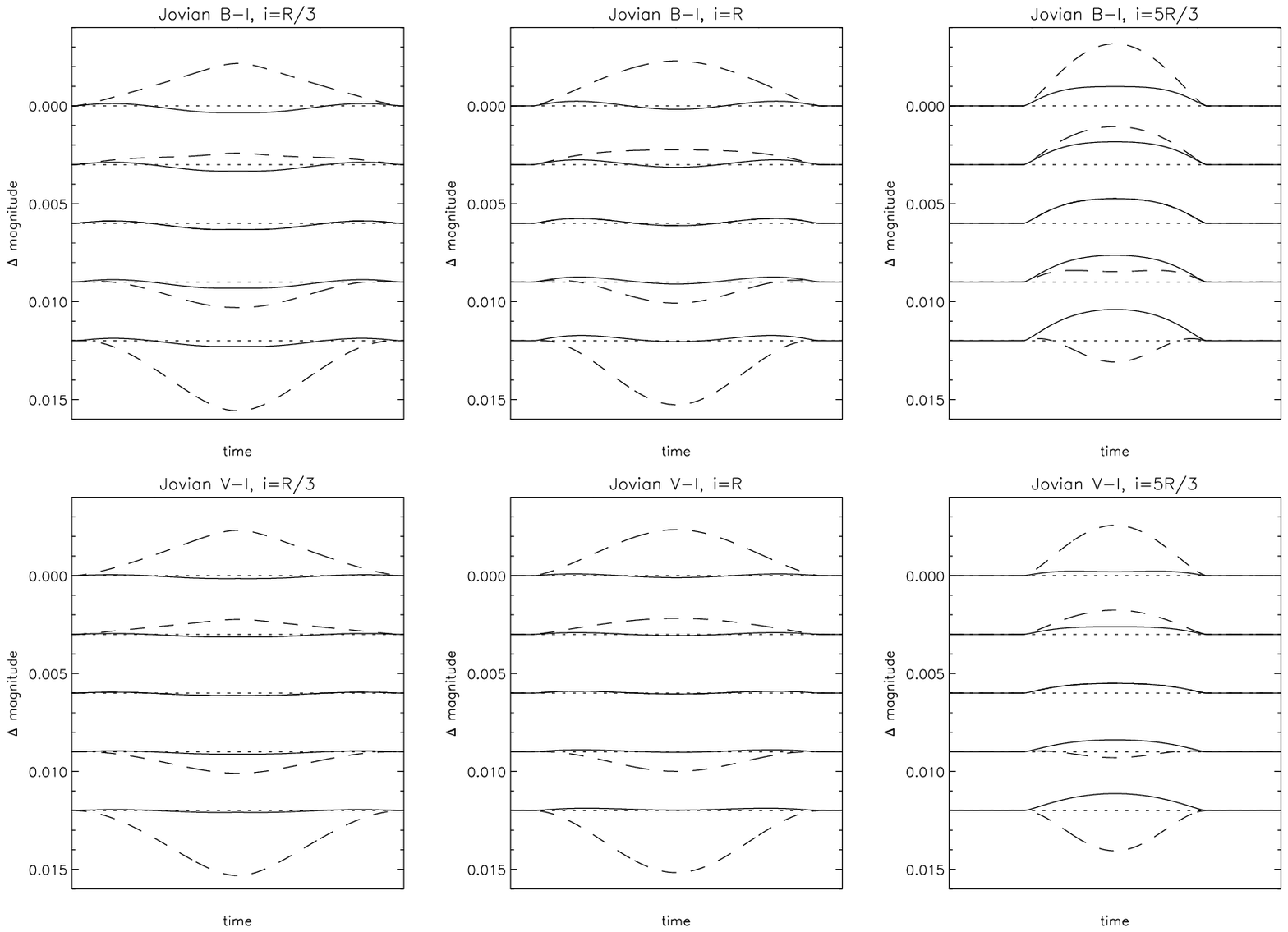}
   \caption{This figure shows the primary and secondary transits for the
case where one of the the eclipsing components is different from
the other two components in the blend for different projected inclinations 
(designed with i) and color differences. The top line in each figure
is color +0.3, followed by +0.1, +0.0, -0.1 and down to -0.3 on the bottom.
The primary eclipse is represented by a dashed line while the secondary is
a solid line.}
\end{figure*}

The corresponding figures for terrestrial-class blends are not shown.
Blended events of this depth are virtually identical in shape to those
shown in Figure 2 for blended Jovian-class events, except that they
are scaled down by a factor of approximately 100. Moreover, terrestrial
exoplanets will very nearly always cross the limb of the
star entirely if they transit at all, as they are much smaller than
stars. Strictly from geometrical arguments, terrestrial planets,
which have a radius approximately 1\% that of a typical star, would 
completely cross the limb of the star approximately 98\% of all
transiting cases. Most likely these extreme events would not be
detected at all, as they will not last very long and would have a
very low amplitude. As such, they will be left out of the blend
analysis for the sake of brevity -- except in the potentially
interesting case of background transiting Jovian exoplanet blends.

Figures 3 explores differences caused by changing the color of
one of the stars in the eclipsing component, showing the primary and
second transits in $\Delta(B-I)$ and $\Delta(V-I)$ for three different
projected inclinations and a variety of color changes. These figures
clearly show that altering the binary pair significantly changes the
primary and secondary
transits. Obviously, an exoplanet only produces one transit per orbit,
which will always look the same. This effect can therefore be used to
differentiate between exoplanets and blends, although color differences
again appear to be the dominant variable. Another potential variable,
the size difference between the stars in the eclipsing components,
was also explored
but found to have very little impact for reasonable (10\%) changes.

\begin{figure*}
   \centering
   \includegraphics{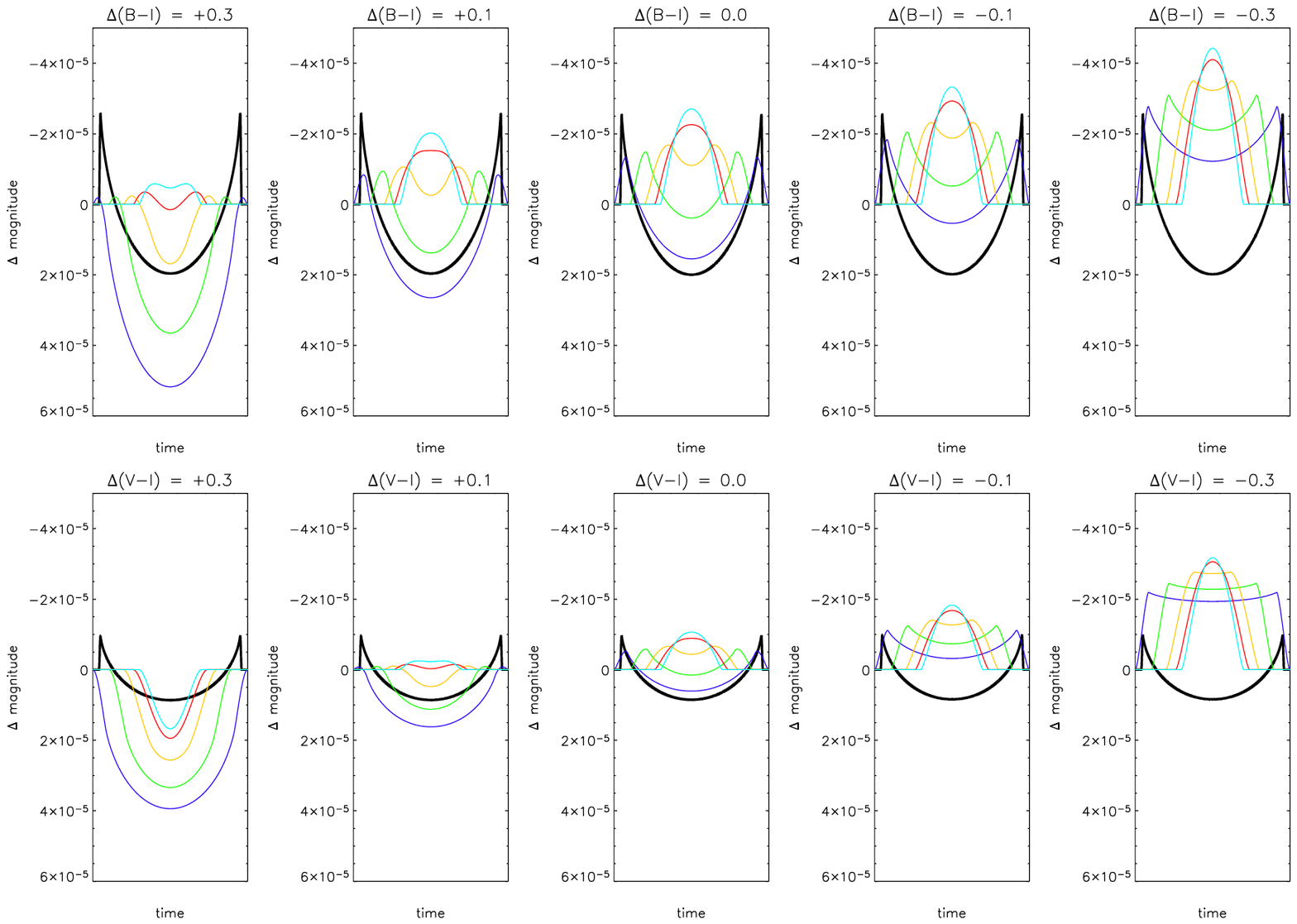}
   \caption{This figure shows the effect of color change on transiting
Jovian exoplanets blended with a constant component to form a terrestrial-class
event. The thick solid line is a central terrestrial transit, while the thin lines
are the blended events with different projected inclinations -- 0 (broadest),
$\frac{2R}{3}$, $0.87 R$, $0.96 R$ and $R$ (narrowest). Central Jovian
blends can resemble terrestrial transits, but even small color changes
between the eclipsing component and the constant component} in the blend
have a strong effect on the blended events.  
\end{figure*}

Another possibility that should be examined is one where a
background transiting Jovian exoplanet is the eclipsing component
of a blend. When combined with a relatively bright constant component,
the depth of the transit of a Jovian exoplanet can be reduced to the
point that it is appears terrestrial in origin. Satellite projects
such as Corot and Kepler will have several tens of
thousands of target stars, generally between magnitudes 10 and 14.
While only a fraction of these target stars will have background stars
of the type that could potentially have exoplanets -- a quick
examination of an OGLE-III bulge field (Udalski et al.
\cite{udalski3}) suggests that only about 1 out of 60 of the stars
between magnitudes 14 and 14.5 will have a background star that is two to
four magnitudes fainter within an arcsecond, much less close enough to make
a true blend -- the possibility remains.

Figure 4 depicts transiting Jovian planet blends for several
projected inclinations and
color differences. Despite the fact that Jovian exoplanetary blends
can produce double-horned profiles, only in the case of a central
Jovian transit will the blended events resemble terrestrial transits --
and then only when the color difference between the constant component 
and the exlipsing component is very small (less than 0.1). Even small color
changes have a strong effect on the shape of the blended events. 

\section{Conclusions}

The color index of transits is a useful discriminator between exoplanets
and blends. Non-grazing exoplanet transits exhibit a double-horned
profile that is distinct from the bell-curves produced by blends.
As the limits of this method are technical (instrumental precision) rather
than intrinsic (asteroseismic oscillations in radial velocity), it can be
used on exoplanet candidates that are both swallower and fainter
than the radial velocity technique currently employed for the task.
Moreover, any color difference between the components of a blend
increases the strength of the blend signature. This, coupled with the
likelihood that the fainter eclipsing component will
be redder than the static component in most cases (either because it is a
foreground M-dwarf pair or because it lies in the background and therefore
subject to more extinction than the static component), means that the
signature from a blend will often be stronger than the signal from an
exoplanet, making them easier to identify.

There are only a few cases where this technique fails. Grazing exoplanet
transits fail to produce distinctive double-horned signatures -- only about
20\% of all Jovian exoplanets and about 2\% of all terrestrial exoplanets
will cause grazing transits. Even in these cases, the color change in $B-I$
is twice that as in $V-I$, a characteristic not likely to be produced by a
blend, as it requires the color differences between in the static component
and the eclipsing component ($\Delta(B-I)$ and $\Delta(V-I)$) to be the
within approximate $0.05$ of each other. This is a relatively unlikely
scenario in the presence of extinction.

For Jovian-class events, it is difficult but not impossible to obtain the
precision needed to observe the difference between blends and exoplanets
from ground. A Jovian exoplanet would produce a signature with a strength
in excess of 1 mmag, which is approximately the limit for ground-based
observations in a single color magnitude (see, for example, Gilliland et al.
\cite{gilliland}). Creating a color from two color
magnitudes decreases the precision, but the relatively long duration of
the events allows for it to be observed many times each transit and their
periodic nature allows for the transit to be observed multiple times.
This makes is possible to build up the necessary statistics to enable a
credible classification.

It is far more difficult to classify terrestrial-class events. Their
expected signature would be around 10 ppm, which is not possible with
any current instrument. The HST observations of the transiting exoplanet
in HD 209458 reached a precision of 110 ppm, for example (Brown et al.
\cite{brown}), using absolute photometry. Instruments designed specifically
for precision photometry can do better. One example of this is the Kepler
mission, which expects a noise level of approximate 74 ppm at the 15
minute level for a G2 star with $m_R = 12$ (Jenkins \& Doyle
\cite{jenkdoyl}. Another, the recently-cancelled MONS mission, was
specifically designed for precision two-color photometry of a single
target star. It expected to reach a noise level of about 54 ppm at the
15 minute level in color intensity ratio -- equivalent to color -- for a
star with $m_V = 5$  (Christensen-Dalsgaard \cite{jcd}). That means that
a precision of 10 ppm could be reached in about 13.7 hours for Kepler in
one color magnitude and about 7.2 hours for MONS in an actual color. As
the duration of an Earth twin transit is about 13 hours, these numbers
are interesting, provided one neglects stellar activity. Detecting
signals in the presence of this sort of non-white noise is
another topic entirely, however.

The value of a transit candidate is limited without the ability to
classify it. In the light of this analysis, it would be very practical
for missions that intend to search for transiting terrestrial planets
to include some way of obtaining color information. Otherwise, a
smaller secondary mission would be required in order to obtain the
necessary observations. It should be noted that such an instrument
would be of tremendous scientific value, not just for following-up
transit candidates but also for any phenomena that involves small
changes in intensity, such as asteroseismology.

\begin{acknowledgements}
 
I would like to thank the Danish Natural Sciences Research Council for
financial support. I would also like to thank Hans Kjeldsen and J\o rgen
Christiansen-Dalsgaard for all the support they have provided me over the
past two years. And lastly I would like to thank Torben Arentoft and
Grzegorz Kopacki for putting up with my constant questions.

\end{acknowledgements}

\end{document}